\documentclass[twocolumn,english,nofootinbib]{revtex4-1}
\usepackage[T1]{fontenc}
\usepackage{textcomp}
\usepackage[latin9]{inputenc}
\usepackage{amsmath}
\usepackage{amssymb}
\usepackage{graphicx}
\usepackage{babel}
\begin{document}
\title{Energy accreted onto a compact star}
\author{Kaiser Arf and Kai Schwenzer}
\address{Istanbul University, Science Faculty, Department of Astronomy and
Space Sciences, Beyazit, 34119, Istanbul, Turkey}
\begin{abstract}
We compute the energy gained by accretion onto a compact star due
to matter falling from the inner edge of a thin accretion disk. We
employ a controlled slow-rotation expansion based on the Hartle-Thorne
metric and find that in neutron stars in general only leading order
rotational corrections to the metric, describing frame dragging, are
sizable. However, for accretion onto a magnetized neutron star, where
the disk roughly extends to the co-rotation radius, even frame dragging
effects are minor. Based on general conservation laws we then derive
simple, equation-of-state-independent expressions for the spin-up
and heating energy, consistently including the relevant rotational,
relativistic and nuclear effects, and find that in observed accreting
millisecond sources the results can strongly deviate from presently
employed estimates.
\end{abstract}
\maketitle

\section{Introduction}

Accretion onto compact objects drives many of the most powerful signals
in the universe. Due to angular momentum conservation the accreted
matter forms an accretion disk and the simplest case, typically realized
if the companion is not too close and the accretion rate not too large,
is a thin disk \citep{Shakura:1972te,Abramowicz:2011xu}, that is
effectively constrained to the central plane and the gained gravitational
energy of material in the disk is efficiently radiated. Although accretion
disks generally can be described in Newtonian approximation as far
as they are sufficiently detached from the accreting compact object,
when the matter eventually falls from the inner edge of the disk towards
a stellar-mass compact object it traverses regions of huge spacetime
curvature and therefore this requires a controlled analysis within
general relativity \citep{1973grav.book.....M,1985fcgr.book.....S,2004sgig.book.....C}.
A key quantity for the emission of compact binaries is the energy
deposited by the accreted matter and its knowledge has the potential
\citep{Arf:2024lgp} to inform us about novel phases of dense matter
\citep{Alford:2019oge} in their interior. In such astrophysical applications,
e.g. \citep{Brown1998,Yakovlev:2002ti,Wijnands2017,Potekhin2019,Potekhin:2023ets},
the accreted energy is often estimated within a simplistic Newtonian
approximation assuming all the energy when matter falls onto the compact
object from infinity is emitted and observed in outburst. In this
work we perform a controlled analysis based on conservation laws and
a consistent slow rotation expansion to determine the rotational and
thermal energy gain per baryon, which improves on this naive estimate
in several ways, taking into account that: (i) the space time around
a compact object is strongly warped and (ii) also rotates owing to
the effect of frame dragging, (iii) matter generally falls only from
the inner edge of an accretion disk but (iv) initially also carries
rotational energy, and (v) the rotational part of the kinetic energy
of the accreted matter spins up the compact object while merely the
non-rotational part heats the surface of a compact star leading to
the astrophysically observable emission. Our results in principle
likewise apply to slowly rotating black holes, where in contrast the
non-rotational energy is lost while only part of the rotational energy
could be extracted subsequently, e.g. via the Blandford-Znajek mechanism
\citep{Blandford:1977ds}, potentially resulting in the observed powerful
non-thermal emission.

\section{Applicability of the slow rotation expansion\label{sec:Applicability-of-the}}

\begin{figure}
\includegraphics[scale=0.65]{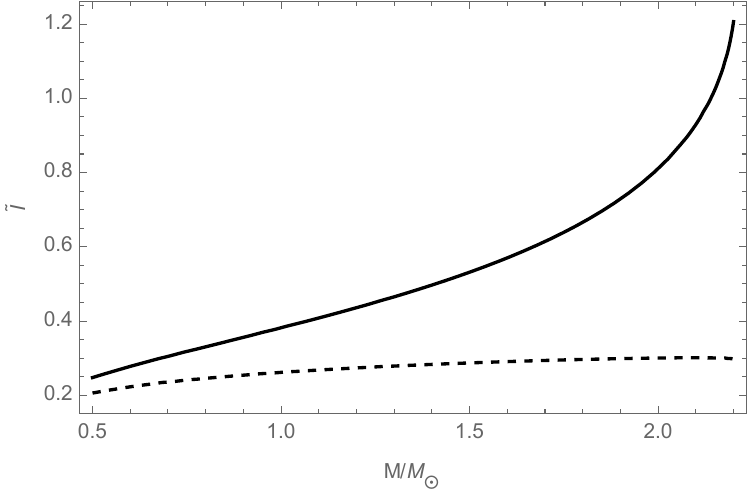}

\caption{\label{fig:moment-of-inertia} The dimensionless moment of inertia
$\tilde{I}=I/\left(MR^{2}\right)$ for an APR equation of state \citep{Akmal:1998cf}
(with maximum mass $\sim2.2M_{\odot})$ when using the relativistic
expression eq. (\ref{eq:integrated-quantities}) below (solid), as
well as its non-relativistic approximation (dashed).}
\end{figure}

While, according to the no hair theorem \citep{PhysRevLett.26.331},
a black hole is completely determined by its conserved quantities,
described in the electrically neutral case by the Kerr metric \citep{Kerr1963},
the spacetime around a compact star is more complicated and depends
to some extend on the structure of the star. Hartle and Thorne \citep{Hartle:1968si}
determined the spacetime metric around a compact object in a controlled
slow rotation expansion

\begin{align}
 & ds^{2}_{\mathrm{Hartle-Thorne}}=\tilde{g}_{tt}dt^{2}+g_{rr}dr^{2}\label{eq:Hartle-Thorne-metric}\\
 & \qquad+g_{\theta\theta}\left(d\theta^{2}+\sin^{2}\theta\left(d\phi-\frac{2J}{r^{3}}dt\right)^{2}\right)+O\!\left(a^{3}\right)\nonumber 
\end{align}
where the metric signature is $\left(-+++\right)$, geometrized units
($G\!=\!c\!=\!1)$ are used and the coefficients are complicated expressions
given below. The Hartle-Thorne metric depends on three parameters
namely the conserved mass $M$ the conserved angular momentum $J$
(respectively the spin parameter $a=J/M$) as well as the mass quadrupole
moment $Q=J^{2}/M+Q_{0}$, where the first term describes the deformation
due to the rotation and $Q_{0}$ the rotation-independent deformation
sustained by forces (e.g. nuclear or magnetic). The latter describes
the particular matter distribution and correspondingly the ``hair''
of a neutron star to leading order in the expansion. It is useful
to express these quantities in terms of dimensionless parameters.
They quantify (the inverse radial position bounded due to) the finite
size of the source ($\chi$), the spin ($\alpha$), and the rotation-independent
deformation $(\vartheta)$ of the compact object. The latter vanishes
for the special case of a Kerr black hole \citep{Kerr1963} and the
size of these parameters can be estimated for neutron stars

\begin{align}
\chi & \equiv\frac{R_{S}}{r}=\frac{2GM}{c^{2}r}\approx\left(0.34\!\pm\!0.03\right)\left(\frac{M}{1.4\,M_{\odot}}\right)\frac{R}{r}\label{eq:expansion-parameters}\\
\alpha & \equiv\frac{a}{R_{S}}=\frac{2\pi c\tilde{I}R^{2}f}{2GM}\approx\left(0.36\!\pm\!0.05\right)\tilde{I}\left(\!\frac{f}{500\,\mathrm{Hz}}\!\right)\!\left(\!\frac{1.4\,M_{\odot}}{M}\!\right)\nonumber \\
\vartheta & \equiv\frac{Q}{MR^{2}_{S}}-\alpha^{2}=\frac{Q_{0}}{MR^{2}_{S}}\lesssim O\!\left(10^{-7}\!-\!10^{-2}\right)\nonumber 
\end{align}
These are by definition smaller than one, while $\chi$ even obeys
the rigorous bound $\chi\leq8/9$ \citep{1983bhwd.book.....S}. Here
it is used that the angular momentum of a slowly rotating star is
$J=I\Omega$, where the moment of inertia \citep{Hartle:1967he},
given below in eq. (\ref{eq:integrated-quantities}), is expressed
as $I=\tilde{I}MR^{2}$ in terms of a dimensionless factor $\tilde{I}\lesssim1$
that describes the particular mass distribution of the neutron star.
In a non-relativistic approximation $\tilde{I}<2/5$ \citep{Alford:2012yn},
but taking into account general relativistic effects it can be larger,
as seen in fig. \ref{fig:moment-of-inertia}.

Note that even though a neutron star spins much slower than a maximally
spinning black hole, the spin parameter can be sizable since a neutron
star is at the same time much larger. For orbits around a neutron
star $r\geq R$. The radius of a neutron star as obtained from TOV
solutions, is (sufficiently away from the mass limit) weakly dependent
on the mass. Moreover, realistic sources cannot be arbitrarily compact
and have to be sufficiently away from their mass limit, taking into
account that during their long lifetime of up to billions of years
accreting sources regularly survived cataclysmic events, like thermonuclear
explosions spreading over their entire surface. The radius of the
two sources PSR J0437--4715 ($M\approx1.418\pm0.037\,M_{\odot},$
$R\approx11.36^{+0.95}_{-0.63}\,\mathrm{km}$) \citep{Choudhury:2024xbk}
and PSR J0740-6620 ($M\approx2.073\pm0.069\,M_{\odot},$ $R\approx12.49^{+1.28}_{-0.88}\,\mathrm{km}$)
\citep{Salmi:2024aum} has been narrowly constrained by NICER observations.
Taking into account that neutron star radii generically decrease with
mass gives within the presently observed mass range of accreting sources
$1.4\,M_{\odot}\lesssim M\lesssim2.1\,M_{\odot}$ the narrow range\footnote{I.e. for this estimate we neglect the theoretical possibility of self-bound
strange stars \citep{Witten:1984rs}, where the mass would increase
with radius, taking into account that previous observational estimates
gave nearly identical masses for the two sources \citep{Miller:2019cac,Miller:2021qha},
and we doubled the resulting uncertainty range, considering that these
studies still involve significant uncertainties \citep{Guver:2025wsj}.} $\bar{R}\approx\left(12.0\pm0.8\right)\,\mathrm{km}$. For (the
rotation-independent part of) the quadrupole moment perpendicular
to the rotation axis of various millisecond pulsars (MSPs) there are
strong bounds $Q_{22}\lesssim O\!\left(10^{30}-10^{35}\,\mathrm{kg\,m^{2}}\right)$
from the non-observation of continuous gravitational waves \citep{LIGOScientific:2025kei},
which leads to the above estimates. The quadrupole moment in accreting
sources could in principle be larger, but taking into account that
the lowest bounds from MSPs strongly limit the rigidity of the crust
of \emph{any} neutron star and that an accreted ocean even reduces
the quadrupole moment, it can be expected that non-axisymmetric deformations
are small in accreting sources as well.

Expressed in terms of these dimensionless parameters the metric coefficients
\citep{Hartle:1968si} read\begin{widetext}

\begin{align}
\tilde{g}_{tt} & =-\left(1-\chi+\frac{1}{2}\chi^{4}\alpha^{2}\right)\left(1+2\left(\frac{1}{2}\chi^{3}\alpha^{2}\left(1+\frac{\chi}{2}\right)+\frac{5}{2}\vartheta Q^{2}_{2}\!\left(\frac{2}{\chi}-1\right)\right)P_{2}\!\left(\cos\!\theta\right)\right)\nonumber \\
g_{\theta\theta} & =r^{2}\left(1+2\left\langle -\frac{1}{2}\chi^{3}\alpha^{2}\left(1+\chi\right)+\frac{5}{2}\vartheta\left(\frac{\chi}{\sqrt{1-\chi}}Q^{1}_{2}\!\left(\frac{2}{\chi}-1\right)-Q^{2}_{2}\!\left(\frac{2}{\chi}-1\right)\right)\right\rangle P_{2}\!\left(\cos\theta\right)\right)\nonumber \\
g_{rr} & =\left(1-\chi+\frac{1}{2}\chi^{4}\alpha^{2}\right)^{-1}\left(1-2\left(\frac{1}{2}\chi^{3}\alpha^{2}\left(1-\frac{5\chi}{2}\right)+\frac{5}{2}\vartheta Q^{2}_{2}\!\left(\frac{2}{\chi}-1\right)\right)P_{2}\!\left(\cos\theta\right)\right)\label{eq:Hartle-Thorne-coefficients}
\end{align}
\end{widetext}

Here $P_{2}\!\left(\cos\theta\right)$ is the corresponding Legendre
polynomial and the arising functions are given by

\begin{align*}
Q^{1}_{2}\!\left(\xi\right) & \equiv\sqrt{\xi^{2}\!-\!1}\left(\!\frac{3\xi^{2}\!-\!2}{\xi^{2}\!-\!1}\!-\!\frac{3}{2}\xi\log\!\left(\frac{\xi\!+\!1}{\xi\!-\!1}\right)\negmedspace\right)\!\xrightarrow[\xi\gg1]{}\!\frac{2}{5\xi^{3}}\!+\!\cdots\\
Q^{2}_{2}\!\left(\xi\right) & \equiv\frac{3}{2}\left(\xi^{2}\!-\!1\right)\log\!\left(\frac{\xi\!+\!1}{\xi\!-\!1}\right)\!-\!\frac{3\xi^{3}\!-\!5\xi}{\xi^{2}\!-\!1}\!\xrightarrow[\xi\gg1]{}\!\frac{8}{5\xi^{3}}\!+\!\cdots
\end{align*}
Here the asymptotic expansion presents an approximation better than
the 10\% level for $\xi>3$, i.e. for $\chi<1/2$ which is realized
in presently observed neutron stars as estimated above, so that

\[
Q^{i}_{2}\!\left(\frac{2}{\chi}\!-\!1\right)\xrightarrow[r\gg R_{s}]{}\frac{\chi^{3}}{4^{2-i}5\left(1\!-\!\frac{\chi}{2}\right)^{3}}
\]
This simple asymptotic dependence allows us to estimate the size of
the different terms in the Hartle-Thorne metric: All rotation terms
in the metric coefficients eq. (\ref{eq:Hartle-Thorne-coefficients})
involve $\alpha^{2}$ and are additionally suppressed by $\chi^{3}$
(although we do not expand in $\chi$). Considering the the fastest
spinning observed millisecond pulsar PSR J1748\textminus 2446ad \citep{Hessels:2006ze}
and the seemingly identical frequency of the fastest low mass x-ray
binary 4U 1820-30 \citep{Jaisawal:2024lps} this gives for observed
sources a combined suppression factor

\[
\alpha^{2}\chi^{3}\lesssim\left(2.0\pm0.8\right)\%\left(\frac{f}{716\,\mathrm{Hz}}\right)^{2}\frac{M}{2.5\,M_{\odot}}\left(\frac{R}{r}\right)^{3}
\]
which is at the percent level outside of any presently observed neutron
star. Similarly the terms involving the quadrupole moment are likewise
additionally suppressed and furthermore all come with at least an
algebraic factor $1/2$ in eqs. (\ref{eq:Hartle-Thorne-coefficients})
so that they are of order

\[
\frac{1}{2}\chi^{3}\vartheta=\left(12\!\pm\!2\right)\%\left(\!\frac{M}{2.5\,M_{\odot}}\!\right)^{3}\!\left(\!\frac{R}{r}\!\right)^{3}\!\frac{Q_{0}}{MR^{2}_{S}}\lesssim O\!\left(10^{-3}\right)
\]
and suppressed by an order of magnitude even for the completely unrealistic
case $Q_{0}/\left(MR^{2}_{S}\right)=O\left(1\right)$---describing
deformations of the order of the Schwarzschild radius. Realistic deformations
should be significantly smaller and taking as a guide the bounds for
millisecond pulsars from gravitational wave searches \citep{LIGOScientific:2025kei}
these corrections should at most be at the per mill level and likely
even much smaller, so that neutron stars have ``nearly no hair''.
Due to all this the metric coefficients describing neutron stars,
up to small corrections at the percent level, strongly simplify

\[
\tilde{g}_{tt}\approx-\left(1-\chi\right)\quad,\quad g_{\theta\theta}\approx r^{2}\quad,\quad g_{rr}\approx\left(1-\chi\right)^{-1}
\]
which is precisely their Schwarzschild form. However, there is still
the term linear in $J$ (and correspondingly in $\alpha$) in eq.
(\ref{eq:Hartle-Thorne-metric}) which is not strongly additionally
suppressed. Therefore for compact stars the Hartle-Thorne metric \citep{Hartle:1968si}
in general reduces to an accuracy at the percent level to the much
simpler Lense-Thirring metric \citep{Lense:1918zz}

\begin{equation}
ds^{2}_{\mathrm{Lense\text{-}Thirring}}=ds^{2}_{\mathrm{Scharzschild}}-\frac{4Ma}{r}\sin^{2}\theta dtd\phi+O\!\left(a^{2},Q\right)\label{eq:slow-rotation-metric}
\end{equation}
where the correction term linear in $a$ describes the effect of frame
dragging. It is worth pointing out that this is a general result that
goes beyond the scope of the present article.

\section{Constants of motion of a particle in a circular orbit\label{sec:Constants-of-motion}}

\begin{figure}
\includegraphics[scale=0.65]{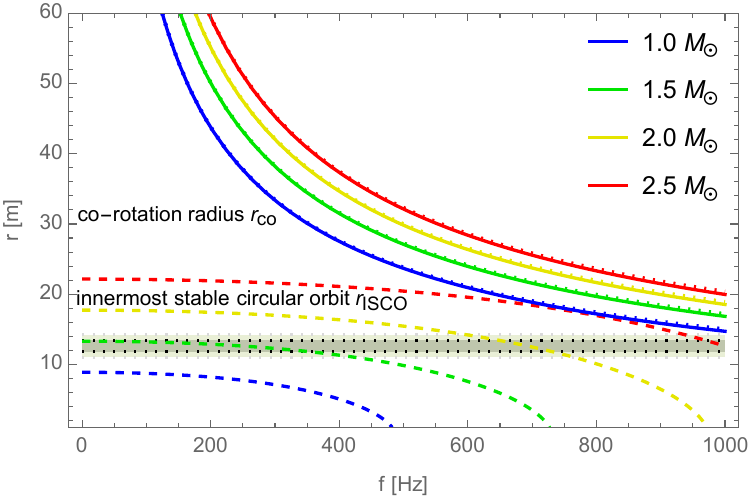}

\caption{\label{fig:Kepler-radius} Innermost stable circular orbit (ISCO,
dashed) and co-rotation radius $r_{\mathrm{co}}$ (solid) for a star
spinning with frequency $f$ in a Hartle-Thorne spacetime, as well
as the standard Kepler law in the Newtonian case (dotted, nearly identical
to the solid lines), for different star masses.These present roughly
the range of the inner disk radius when accretion is possible and
the dotted horizontal band shows the current uncertainty for the neutron
star radius from observed sources.}
\end{figure}

Owing to angular momentum conservation matter accreted onto a compact
object from a companion in a sufficiently distant orbit forms an accretion
disk, where angular momentum is slowly transferred via viscous forces
until the matter can eventually be accreted \citep{Shakura:1972te}.
In this process any initial asymmetry is eventually damped away so
that the matter in the disk to good approximation rotates in circular
orbits in the central plane. In the Newtonian case the motion of a
particle in any circular orbit obeys Kepler's third law. Considering
circular geodesics in the central plane, in both the Kerr \citep{Kerr1963}
and the above approximate Lense-Thirring \citep{Lense:1918zz} metric
a generalized version is obtained \citep{Bardeen:1972fi}

\begin{equation}
\omega=\pm\frac{M^{\frac{1}{2}}}{r^{\frac{3}{2}}\pm aM^{\frac{1}{2}}}\xrightarrow[a=0]{}\pm\frac{M^{\frac{1}{2}}}{r^{\frac{3}{2}}}\label{eq:general-Kepler-law}
\end{equation}
where the frequencies in the different directions are distinct due
to the effect of frame dragging. The expression reduces to Kepler's
non-relativistic law if the central object is non-rotating, showing
that the latter also holds in the relativistic Schwarzschild case
\citep{1973grav.book.....M}. In the absence of other forces orbits
are stable outside of the Innermost Stable Circular Orbit (ISCO),
which in a Schwarzschild metric is at $r_{\mathrm{ISCO}}=3\,R_{S}$
\citep{1949ZhETF..19..951K} and with increasing spin it moves closer
to the star \citep{Bardeen:1972fi}. However, viscous heating in the
disks easily turns atomic matter into a plasma and compact objects
generally have large magnetic fields which complicate the situation
since there are other forces than gravity that determine the stability
of orbits. The most important one is the Lorentz force due to the
magnetic field of the star, but e.g. also the contribution from a
plasma in the magnetosphere or the ram pressure can have a (smaller)
impact, if e.g. the wind from the compact object is strong. Within
the light cylinder the magnetic field rotates rigidly with the compact
object. Owing to the Lorentz force charged particles gyrate narrowly
around the field lines so that they are effectively forced to move
along them when the field is strong enough \citep{Goldreich:1969sb}.
In this case roughly the co-rotation radius, where a particle rotates
as fast as the star, $\omega=\Omega$, is the only (unstable) orbit
and represents the limiting radius within which matter is accreted
onto the compact object\footnote{In contrast, matter rotating outside of the co-rotation radius experiences
fictitious outward forces that are overturned by the interactions
with inflowing matter, though.}. The co-rotation radius as a function of the frequency reads

\begin{equation}
\frac{r_{\mathrm{co}}}{R_{s}}=\frac{\left(1-a\Omega\right)^{\frac{2}{3}}}{2\left(M\Omega\right)^{\frac{2}{3}}}\approx4.1\left(1-a\Omega\right)^{\frac{2}{3}}\!\left(\!\frac{M}{1.4\,M_{\odot}}\!\right)^{-\frac{2}{3}}\!\left(\!\frac{f}{\mathrm{kHz}}\!\right)^{-\frac{2}{3}}\label{eq:co-rotation-radius}
\end{equation}
where the arising correction factor due to the star's rotation is
for neutron stars bounded by

\begin{equation}
a\Omega=\tilde{I}\left(R\Omega\right)^{2}\lesssim\left(3.2\pm0.4\right)\!\%\,\tilde{I}\left(\frac{f}{\mathrm{716\,Hz}}\right)^{2}\label{eq:rotation-parameter}
\end{equation}
in terms of the above parametrization of the moment of inertia, and
the radius of the star $R$ has again been estimated from \citep{Choudhury:2024xbk,Salmi:2024aum}.
Therefore, for presently observed sources with frequencies $f\lesssim716\,\mathrm{Hz}$,
the effect of the rotational corrections on the co-rotation radius
is smaller than uncertainties of typical neutron star observables
at present. This is seen in fig. \ref{fig:Kepler-radius} which shows
the obvious statement that the disk in fast rotating sources extends
much closer to the compact object than in slowly rotating sources.
The leading order Hartle-Thorne/Lense-Thirring result for the co-rotation
radius including the effect of frame dragging due to eq. (\ref{eq:slow-rotation-metric})
(solid) is compared to the non-relativistic Kepler result (dotted)
and as can be seen the deviations are indeed small. This can be traced
back to the fact that even though the leading frame dragging term
in the metric is linear in $\alpha\sim a\sim\Omega$, the arising
correction factor $a\Omega$ in the co-rotating radius eq. (\ref{eq:co-rotation-radius})
is quadratic in $\Omega$ and therefore roughly of the same order
as the corrections neglected in the metric above. As can be seen the
co-rotation radius is clearly above the radius of observed neutron
stars denoted by the gray horizontal band, so a thin plasma disk is
always detached from the star. The dashed curves show the ISCO, and
as can be seen for light stars it is always inside of the star radius,
while for heavy stars it is inside at sufficiently high spin frequency.
In such cases a hypothetical neutral component of the disk could correspondingly
extend down to the surface and otherwise at least considerably closer
to the star.

\begin{figure}
\includegraphics[scale=0.65]{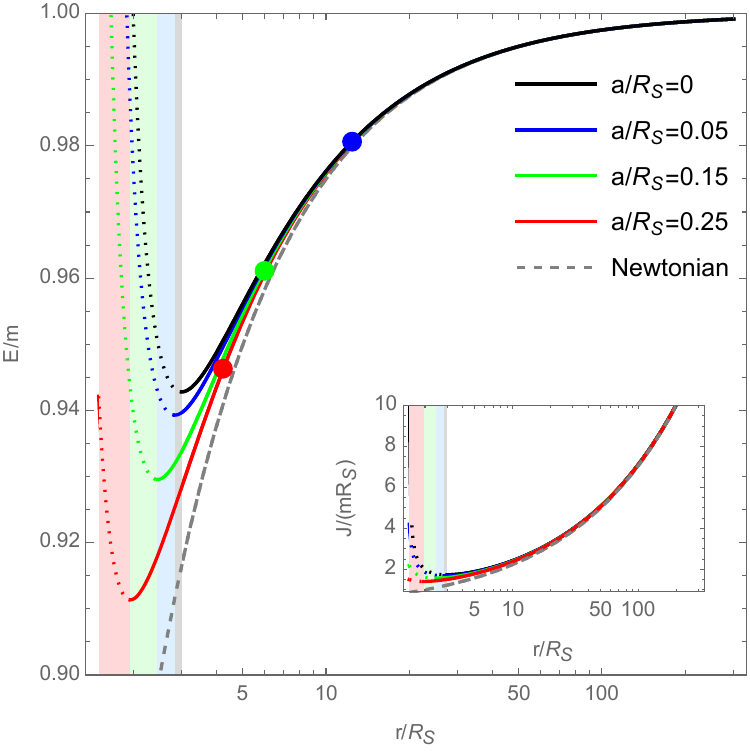}

\caption{\label{fig:conserved-energy} The conserved energy of a mass in a
circular orbit in a Hartle-Thorne metric assuming $M=1.4\,M_{\odot}$
for different spin parameters (solid), where a value of zero reduces
to the Schwarzschild metric, as well as in the Newtonian limit (dashed).
The dots show for comparison the co-rotation radius and the inset
shows the analogous plot for the conserved angular momentum.}
\end{figure}

Any explicitly time-independent metric, like the Hartle-Thorne metric,
has a timelike Killing vector $\partial/\partial t$ that leads to
a conserved energy $E$, and the axial symmetry yields the second
Killing vector $\partial/\partial\phi$ leading to a conserved angular
momentum $J$ \citep{2004sgig.book.....C}. In the coordinates specified
by eq. (\ref{eq:Hartle-Thorne-metric}) the Killing vectors have the
explicit form $k^{\mu}=\left(1,0,0,0\right)$ and $\mu^{\mu}=\left(0,0,0,1\right)$
which yields for a particle of mass $m$ orbiting in the central plane

\begin{align}
E\!\left(r,\omega,a\right) & =-mk^{\mu}u_{\mu}\xrightarrow[\theta=\pi/2]{}m\left(1-\frac{R_{S}}{r}\left(1-a\omega\right)\right)\frac{dt}{d\tau}\label{eq:conserved-energy}\\
J\!\left(r,\omega,a\right) & =m\mu^{\mu}u_{\mu}\xrightarrow[\theta=\pi/2]{}m\left(r^{2}\omega-\frac{R_{S}a}{r}\right)\frac{dt}{d\tau}\label{eq:conserved-angular-momentum}
\end{align}
The arising derivative obtained from eq. (\ref{eq:Hartle-Thorne-metric})
yields $dt/d\tau=\left(1-\frac{R_{S}}{r}\left(1-2a\omega\right)-r^{2}\omega^{2}\right)^{-1/2}$
and using the generalized Kepler relation eq. (\ref{eq:general-Kepler-law})
gives for a particle rotating in a stable orbit with an angular speed
$\omega$

\begin{align}
\frac{E\!\left(\omega\right)}{m} & =\frac{1-2\left(M\omega\right)^{2/3}\left(1-a\omega\right)^{\frac{1}{3}}}{\sqrt{1-\frac{\left(M\omega\right)^{2/3}}{\left(1-a\omega\right)^{\frac{2}{3}}}\left(3\left(1-2a\omega\right)+a^{2}\omega^{2}\right)}}\label{eq:Kepler-energy}\\
\frac{J\!\left(\omega\right)}{m} & =\frac{\frac{M^{\frac{2}{3}}}{\omega^{\frac{1}{3}}\left(1-a\omega\right)^{\frac{2}{3}}}\left(1-4a\omega+a^{2}\omega^{2}\right)}{\sqrt{1-\frac{\left(M\omega\right)^{2/3}}{\left(1-a\omega\right)^{\frac{2}{3}}}\left(3\left(1-2a\omega\right)+a^{2}\omega^{2}\right)}}\label{eq:Kepler-angular-momentum}
\end{align}
The conserved energy and the conserved angular momentum as a function
of radius are shown in fig. \ref{fig:conserved-energy} for different
neutron star spins. While general orbits are determined by an effective
potential \citep{1973grav.book.....M,2004sgig.book.....C} for a circular
orbit in the equatorial plane it reduces to the conserved energy.
In contrast to the Newtonian case in relativity the lowest energy
is obtained at the ISCO which moves inwards for faster spinning sources.
I.e. if magnetic forces are absent and an accretion disk would extend
all the way to the ISCO, the frame dragging corrections could have
a sizable impact on the accretion. However, if magnetic fields are
important and determine the inner edge of a plasma disk, it should
approximately be determined by the co-rotation radius shown by the
dots, which is at significantly larger distances (see also eq. (\ref{eq:co-rotation-radius})
and fig. \ref{fig:Kepler-radius}) where frame dragging is already
minor. Since for faster rotation both the ISCO and the co-rotation
radius move inwards the difference between the corresponding conserved
energies is interestingly only rather weakly dependent on the spin
frequency.

\section{Accretion heating and spin-up\label{sec:Accretion-heating}}

An accretion disk consists of a viscous fluid that also carries thermal
energy. Spectral studies of transiently accreting neutron stars, see
e.g. \citep{Ibragimov:2009js}, show that the temperature of the inner
disk is typically below a keV so that the thermal energy per particle
is of order keV, while, as seen below, the kinetic energy per particle
gained from accretion is of order 100 MeV so that the initial thermal
energy can safely be neglected when computing the energy transferred
to the star, in line with the considered approximation of a cold,
thin disk. Even the rotational energy at the edge of the disk is of
order MeV so that individual particles forming the disk---instead
of entire fluid elements---can be considered to be in Kepler orbits
around the compact object. The conserved quantities of a particle
in a circular orbit eqs. (\ref{eq:Kepler-energy}) and (\ref{eq:Kepler-angular-momentum}),
respectively fig. \ref{fig:conserved-energy} evaluated at the inner
radius of an accretion disk $r_{d}$ present then the initial conditions
a particle has when it falls from a stable orbit of corresponding
angular speed $\omega$ towards the star. Although the trajectory
along a magnetic field line is not a geodesic, magnetic forces merely
deflect the particle, so that the conserved energy stays constant
as the particle descends. As soon as the particle smashes onto the
surface the particle merges with the star to form a slightly more
massive and more compact object that is spun up and heated in the
process. Although the details are complicated\footnote{E.g. a particle could already loose most of its energy in an extended,
thick plasma above the magnetic pole where the observed pulsed power
law x-ray emission \citep{Ibragimov:2009js} seems to stem from. Yet,
the latter is bound to and rotates with the star and correspondingly
can be considered as part of it.} assuming that only a small fraction of accreted matter is ejected
and radiative losses (most notably due to synchrotron emission) are
small, the total energy and angular momentum is approximately conserved
in the process and therefore it can be described by a perfectly inelastic
collision. For any quantity $Q$ describing the star, we define the
change from the initial state $Q_{i}$ to the final state $Q_{f}$
by the difference $\Delta Q\equiv Q_{f}-Q_{i}$. In case of conserved
quantities like energy and momentum this describes their conservation
during the perfectly inelastic collision, i.e. in this case the difference
is just the energy respectively angular momentum of the initial particle
$\Delta E=E\!\left(\omega\right)$ and $\Delta J=J\!\left(\omega\right).$

A static static stellar configuration is determined by its Equation
of State (EoS) and central pressure $p_{0}$, and the final state
of a star with a particle more has a by $\Delta p_{0}$ larger central
pressure than the initial star before the collision. Solving the
Oppenheimer-Volkov (OV) equations \citep{Oppenheimer1939} then yields
a slightly different pressure profile that can be used to compute
astrophysical observables. These are always macroscopic averages over
the entire star, taking into account that due to their tiny size and
huge distance these are inherently point sources. Slowly rotating
stars have been considered by Hartle \citep{Hartle:1967he} and it
was found that structural deformations of the star only enter at quadratic
order in the rotation parameter $a$, while, as discussed, frame dragging
effects already enter at linear order. Since the angular momentum
involves another power of $a$ these rotational corrections of the
moment of inertia or other star properties enter at least at quadratic
order which is small according to eqs. (\ref{eq:expansion-parameters}).
While the result for the moment of inertia has been given in a concise
form in \citep{Hartle:1967he}, using the OV equations it can to leading
order be written in a conventional form in terms of the Schwarzschild
interior metric

\[
ds^{2}_{\mathrm{Schwarzschild}}=-e^{2\phi\!\left(r\right)}dt^{2}+\frac{dr^{2}}{1\!-\!\frac{2m\!\left(r\right)}{r}}+r^{2}\negmedspace\left(d\theta^{2}+\sin\!\theta^{2}d\varphi^{2}\right)
\]
Assume the mass, particle number and moment of inertia\footnote{The form for the moment of inertia given in \citep{Hartle:1967he}
can be transformed into the more conventional form given here using
the OV equations.} \citep{Hartle:1967he}

\begin{align}
 & N\!\left(p_{0}\right)\!=\!4\pi\negthickspace\int^{R}_{0}\negthickspace dr\,r^{2}\frac{n\!\left(p_{p_{0}}\negmedspace\left(r\right)\right)}{\sqrt{1-\frac{2m\left(r\right)}{r}}}+O\!\left(a\right),\nonumber \\
 & M\!\left(p_{0}\right)\!=\!4\pi\negthickspace\int^{R}_{0}\negthickspace dr\,r^{2}\rho\!\left(p_{p_{0}}\negmedspace\left(r\right)\right)+O\!\left(a\right),\label{eq:integrated-quantities}\\
 & I\!\left(p_{0}\right)=\frac{8\pi}{3}\int^{R}_{0}dr\,r^{4}\frac{\rho\!\left(p_{p_{0}}\negmedspace\left(r\right)\right)+p_{p_{0}}\negmedspace\left(r\right)}{\sqrt{1-\frac{2m\left(r\right)}{r}}}e^{-\phi\!\left(r\right)}\frac{\tilde{\omega}\left(r\right)}{\Omega}+O\!\left(a\right)\nonumber 
\end{align}
 are obtained by an explicit numerical OV computation as a function
of the central pressure, one then gets explicitly for the changes
of these quantities

\begin{align*}
 & \Delta N=1\:\Rightarrow\:\Delta M=\left.\frac{dM}{dN}\right|_{N_{i}},\:\Delta I=\left.\frac{dI}{dN}\right|_{N_{i}}
\end{align*}
In a slow rotation expansion all macroscopic physical quantities separate
into a non-rotating contribution and an additive rotational correction.
Similarly in neutron stars where the temperature is always tiny compared
to the chemical potential $T/\mu\ll1$, also this ratio presents an
expansion parameter and so quantities likewise standardly separate
into a zero temperature term (determined by the OV equations) and
an additive temperature correction. The total angular momentum and
energy of the star before and after the collision are therefore given
by \citep{Hartle:1967he,Stergioulas:2003yp}

\begin{align*}
J_{i/f} & =I_{i/f}\Omega_{i/f}\;,\;E_{\mathrm{rot},i/f}=\frac{J^{2}_{i/f}}{2I_{i/f}}\\
E_{i/f} & =M_{i/f}+E_{\mathrm{rot},i/f}+E_{\mathrm{heat},i/f}
\end{align*}
so that momentum conservation gives (using $\Omega_{i}=\Omega)$
\begin{align}
\Delta E_{\mathrm{rot}} & =\Omega J\!\left(\omega\right)-\frac{\Omega^{2}}{2}\frac{dI}{dN}+O\!\left(\Delta^{2}\right)\label{eq:formal-spinup-energy}
\end{align}
and energy conservation yields accordingly for the heating energy
per particle due to accretion from a cold, thin disk 

\begin{align}
\Delta E_{\mathrm{heat}} & =E\!\left(\omega\right)-\Omega J\!\left(\omega\right)-\left(\frac{dM}{dN}-\frac{\Omega^{2}}{2}\frac{dI}{dN}\right)+O\!\left(\Delta^{2}\right)\label{eq:formal-heating-energy}
\end{align}
where $N$, $M$, and $I$ depend on the EoS and generally have to
be determined by a numerical OV solution \citep{Oppenheimer1939}. 

The above expressions seem to require an explicit numerical OV solution
based on a fixed EoS. However, it turns out that they can be evaluated
completely analytically and are actually generally independent of
the EoS. To see this we start with the definitions eq. (\ref{eq:integrated-quantities})
and take into account that due to the one-to-one relation these quantities
can likewise be expressed as monotonic functions of the radius. Thereby
using eqs. (\ref{eq:integrated-quantities}) the arising EoS-dependent
derivative can be rewritten as 

\begin{align}
 & \left.\frac{\partial M}{\partial N}\right|_{N_{i}}=\left.\frac{\partial M}{\partial R}\right|_{R_{i}}\left.\frac{\partial R}{\partial N}\right|_{N_{i}}\label{eq:mass-change}\\
 & =\left(\frac{\left.4\pi R^{2}n_{i}\left(R\right)\left(1-\frac{R_{S}}{R}\right)^{-\frac{1}{2}}\right|_{R_{i}}}{\left.4\pi R^{2}\rho_{i}\left(R\right)\right|_{R_{i}}}\right)^{-1}=\mu\sqrt{1-\frac{R_{S}}{R_{i}}}\nonumber 
\end{align}
where the fundamental theorem of calculus was used in the second line
and we introduced the ``reduced mass'' per particle of matter at
the surface (dropping the label $i$)

\begin{equation}
\mu\equiv\lim_{r\to R}\frac{\rho\!\left(r\right)}{n\!\left(r\right)}\label{eq:reduced-mass}
\end{equation}
This constant is determined by nuclear physics and generally rather
close to the baryon mass, as discussed below. I.e. up to this small
deviation the rate $\partial M/\partial N$ is simply the conserved
energy eq. (\ref{eq:conserved-energy}) of a (non-rotating) particle
at the radius of the star $dM/dN=\mu/m\left.E\!\left(R\right)\right|_{\Omega=0}$.
Analogously the change of the moment of inertia can be rewritten in
the form
\begin{align}
\left.\frac{\partial I}{\partial N}\right|_{N_{i}} & =\left.\frac{\partial I}{\partial R}\right|_{R_{i}}\left.\frac{\partial R}{\partial N}\right|_{N_{i}}=\frac{2\mu R^{2}_{i}}{3\sqrt{1-\frac{R_{s}}{R_{i}}}}\label{eq:moi-change}
\end{align}
Inserting the above expressions eqs. (\ref{eq:Kepler-energy}), (\ref{eq:Kepler-angular-momentum}),
(\ref{eq:mass-change}) and (\ref{eq:moi-change}) yields then the
general, explicit result for the spin-up and heating energy due to
accretion of a particle of mass $m$ from a thin disk

\begin{align}
 & \Delta E_{\mathrm{rot}}=\frac{m\frac{R_{S}}{2r_{d}}\left(1-4\omega a+\omega^{2}a^{2}\right)\Omega}{\sqrt{1-\frac{R_{S}}{2r_{d}}\left(3-6\omega a+\omega^{2}a^{2}\right)}}-\frac{\mu R^{2}\Omega^{2}}{3\sqrt{1-\frac{R_{s}}{R}}}+O\!\left(a^{2}\right)\label{eq:general-spinup-energy}\\
 & \Delta E_{\mathrm{heat}}=m\frac{1-\frac{R_{S}}{2r_{d}}\left(2+\frac{\Omega}{\omega}-2\left(\omega+2\Omega\right)a+\omega\Omega a^{2}\right)}{\sqrt{1-\frac{R_{S}}{2r_{d}}\left(3-6\omega a+\omega^{2}a^{2}\right)}}\nonumber \\
 & \qquad-\mu\sqrt{1-\frac{R_{S}}{R}}+\frac{\mu R^{2}\Omega^{2}}{3\sqrt{1-\frac{R_{s}}{R}}}+O\!\left(a^{2}\right)\label{eq:general-heating-energy}
\end{align}
Aside from the basic nuclear physics input encoded in the reduced
mass $\mu$ eq. (\ref{eq:reduced-mass}) and the disk radius $r_{d}$
respectively orbital angular speed $\omega$, which are related by
eq. (\ref{eq:general-Kepler-law}) and are set by other forces, these
simple expressions are entirely determined by gravity. The formal
singularity of the expression at $a=0$ and $r=\frac{3}{2}R_{S}$
corresponds to the unstable photon orbit in a Schwarzschild metric
\citep{1949ZhETF..19..951K,1973grav.book.....M} and reflects the
fact that the slow rotation expansion is an expansion around the Schwarzschild
case. However, this singularity is irrelevant since this radius is
effectively always within the star and the disk radius for massive
particles cannot be smaller than the ISCO which, as seen in fig. \ref{fig:Kepler-radius},
is always at larger values anyway. These expressions are valid as
long as the disk stays sufficiently thin. If the radiative cooling
becomes inefficient so that the disk heats up strongly, the disk becomes
thick and the heating energy of the stars will depend on the detailed
geometry of the disk \citep{2002apa..book.....F}, and in the extreme
case where the star is surrounded by a bulge accretion can even become
spherical so that our results do not apply.

It might seem that since we neglected terms of order $\alpha^{2}$
in the Hartle-Thorne expansion, it would be consistent to do the same
in a Taylor series of eqs. (\ref{eq:general-spinup-energy}) and (\ref{eq:general-heating-energy}).
Yet, in the metric coefficients above the leading terms are one, so
that the corrections are indeed at the percent level and negligible
compared to current astrophysical uncertainties. The rotational energy
in contrast is inherently of higher order in $\alpha$ so that one
cannot truncate the expansion at low order. Similarly, the heating
energy involves a difference of the first two terms which are both
of order $m$, and due to this cancellation the impact of the rotational
terms is larger than eq. (\ref{eq:rotation-parameter}) naively suggests.
Therefore, an expansion of these expression to linear order is insufficient
and seemingly small terms, like the last term in these equations,
describing the change of the moment of inertia of the star, can be
relevant. Instead of expanding to the appropriate higher order it
is then more efficient to use the rather concise expressions eqs.
(\ref{eq:general-spinup-energy}) and (\ref{eq:general-heating-energy}).

\begin{figure}
\includegraphics[scale=0.6]{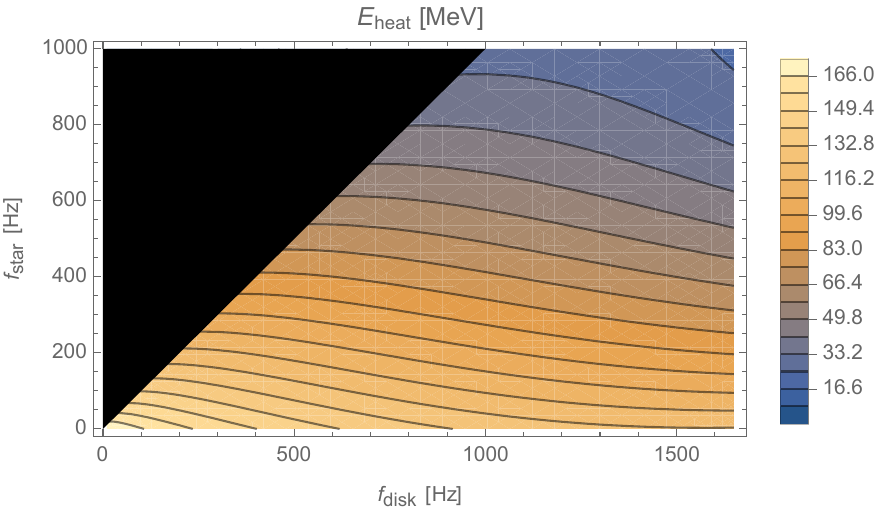}

\caption{\label{fig:general-heating-energy} The heating energy per particle
deposited by matter falling from an inner disk radius rotating at
$f_{\mathrm{disk}}$ onto a $M=1.4\,M_{\odot}$ star rotating with
$f_{\mathrm{star}}$. The diagonal presents the co-rotating case above
which accretion is generally not possible (black region) and the maximum
of the range for the disk frequency corresponds to an orbit at the
radius of the star.}
\end{figure}

The general solution for the heating energy per particle eq. (\ref{eq:general-heating-energy})
due to accretion onto a $M=1.4\,M_{\odot}$ star is shown as a function
of the stellar rotation frequency $f_{\mathrm{star}}=f=\Omega/\left(2\pi\right)$
and of the inner disk frequency $f_{\mathrm{disk}}=\omega/\left(2\pi\right)$
in fig. \ref{fig:general-heating-energy}, using $\mu=m$, see the
discussion below. As can seen the maximum heating energy is obtained
for the co-rotating case $f_{\mathrm{disk}}=f_{\mathrm{star}}$, while
for larger inner disk frequencies, corresponding to smaller radii,
the energy decreases. As is clear from fig. \ref{fig:Kepler-radius},
the minimum possible energy is realized if the disk extends all the
way to the ISCO and this minimum can even be zero if the ISCO lies
inside of the star and the disk correspondingly extends all the way
to the stellar surface. Heavier stars have deeper gravitational potential
wells and correspondingly yield larger accretion energies, as discussed
below, but the generic form of the figure is similar.

As discussed, due to the presence of strong magnetic fields, charged
particles are generally forced to co-rotate with the magnetic field
lines that rotate within the light cylinder approximately rigidly
with the star. If matter falls from the co-rotation radius, the result
for the heating energy eq. (\ref{eq:general-heating-energy}) simplifies
since the numerator and the denominator of the term in the first line
partly cancel

\begin{align}
\Delta E_{\mathrm{heat}} & \xrightarrow[\omega=\Omega]{\mathrm{co-rotation}}m\sqrt{1-\frac{3R_{S}}{2r_{\mathrm{co}}}\left(1-2\Omega a\right)}-\mu\sqrt{1-\frac{R_{S}}{R}}\nonumber \\
 & \qquad+\frac{\mu R^{2}\Omega^{2}}{3\sqrt{1-\frac{R_{s}}{R}}}+O\!\left(a^{2}\right)\nonumber \\
 & \xrightarrow[a=0]{\mathrm{\mathrm{Schwarzschild}}}m\sqrt{1-\frac{3R_{S}}{2r_{\mathrm{co}}}}-\mu\sqrt{1-\frac{R_{S}}{R}}\nonumber \\
 & \xrightarrow[R_{S}\ll R]{\mathrm{Newtonian}}M\left(\frac{\mu}{R}-\frac{3m}{2r_{\mathrm{co}}}\right)\xrightarrow[R\ll r_{\mathrm{co}}]{\mathrm{naive}}\frac{M\mu}{R}\label{eq:heating-energy-corotation}
\end{align}
This result reduces to the Schwarzschild expression for a static
star, to the Newtonian expression in the non-relativistic limit, as
well as to the naive Newtonian gravitational potential---yet with
the reduced mass eq. (\ref{eq:reduced-mass})---when matter falls
all the way from infinity. All these expressions can also be obtained
by a direct computation using the Schwarzschild metric respectively
Newton's laws. While the Newtonian limit is appropriate for white
dwarfs, for neutron stars relativistic corrections are required and
particularly for millisecond sources it is crucial to take into account
that matter falls only from the inner disk radius.

\begin{figure}
\includegraphics[scale=0.65]{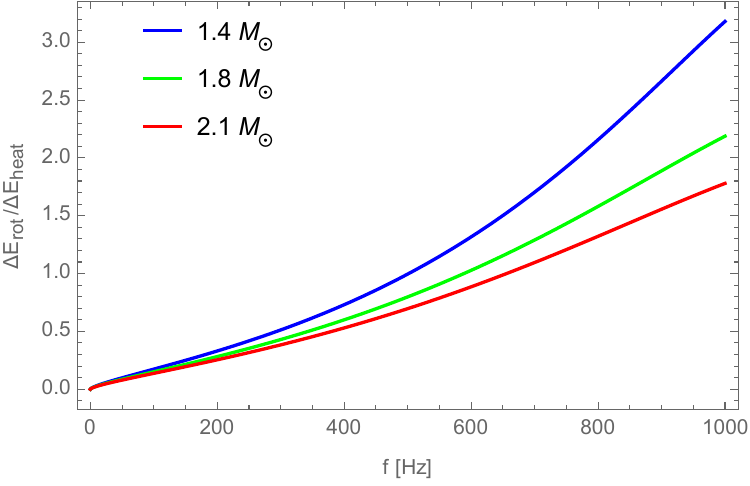}

\caption{\label{fig:rotation-heating-ratio} Spin-up vs. heating energy in
the co-rotating case for different stellar masses.}
\end{figure}

The relative distribution of the accreted energy into spinning up
and heating the star for accretion from the co-rotation radius is
shown in fig. \ref{fig:rotation-heating-ratio}. While at low frequencies
nearly the entire energy heats the star, in the fastest observed millisecond
sources most of the accreted energy ends up spinning up the star.
This can be appreciated by recalling that while the stellar angular
momentum is linear in the spin frequency, the rotational energy is
quadratic. Correspondingly in a faster spinning stars a given angular
momentum increase requires an increasingly larger rotational energy
increase.

\section{Application to different classes of compact stars}

Our general results eqs. (\ref{eq:general-spinup-energy}) and (\ref{eq:general-heating-energy})
(as well as the various limits, where applicable) are completely determined
by general relativity and are valid for any compact star. However,
the reduced mass $\mu$ involves nuclear physics input and needs to
be considered individually for the different cases of interest. In
case of a neutron star (NS) or white dwarf (WD), matter at the surface
becomes non-degenerate so that $p\!\left(R\right)=0$, implicitly
defining $R$, and the energy density is just the mass density. 

The outermost layer of a catalyzed neutron star is ordinary iron with
a binding energy per nucleon of $B_{\mathrm{Fe}}\approx8.8\,\mathrm{MeV}$.
Therefore the arising reduced mass is actually equation of state independent
and simply gives

\begin{align*}
\mu^{\left(\mathrm{Fe}\right)}_{\mathrm{NS}} & =\frac{m_{\mathrm{Fe}}}{A_{\mathrm{Fe}}}=\frac{Z_{\mathrm{Fe}}\left(m_{p}+m_{e}\right)+\left(A_{\mathrm{Fe}}-Z_{\mathrm{Fe}}\right)m_{n}-A_{\mathrm{Fe}}B_{\mathrm{Fe}}}{A_{\mathrm{Fe}}}\\
 & \approx930.5\,\mathrm{MeV}\approx0.991m_{\mathrm{H}}
\end{align*}
However accreting neutron stars are generally not catalyzed and instead
have, analogously to white dwarfs, an ocean of helium and generally
also hydrogen. These can burn explosively, but generally it can be
expected that there is a layer of hydrogen on top in which case $\mu^{\left(\mathrm{H}\right)}_{\mathrm{\mathrm{NS/WD}}}=m_{\mathrm{H}}\approx938.8\,\mathrm{MeV}$.
Merely right after (the probably unrealistic case of) an ideal burst
that burned all the hydrogen, using the binding energy of helium $B_{\mathrm{He}}\approx7.1\,\mathrm{MeV}$,
one would instead have

\begin{align*}
\mu^{\left(\mathrm{He}\right)}_{\mathrm{NS/WD}} & =\frac{m_{\mathrm{He}}}{A_{\mathrm{He}}}\approx932.1\,\mathrm{MeV}\approx0.993m_{\mathrm{H}}
\end{align*}
Since only the surface composition is relevant the same applies to
any hybrid star with an exotic core \citep{Alford:2019oge}. It would
even directly apply to a strange star, made of strange quark matter,
presenting the absolute ground state if the strange matter hypothesis
\citep{Witten:1984rs} is realized, as long as the star has an ordinary
nuclear crust elevated by electrostatic forces and the crust and core
are rotationally coupled. If the suspended crust can rotate approximately
independently from the core, the rather different mass, angular momentum
and moment of inertia of the crust would be relevant in the above
conservation equations, but since only differences between the initial
and final state enter this would not affect the results, as eqs. (\ref{eq:general-spinup-energy})
and (\ref{eq:general-heating-energy}) explicitly show. The same holds
for the presence of rotating superfluids in the interior of an ordinary
neutron star, that have been associated with observed pulsar glitches
\citep{1983bhwd.book.....S}. Correspondingly, the uncertainty stemming
from the composition of any compact object with an ordinary nuclear
surface is negligible and one can approximate $\mu\approx m$. 

In contrast the case of a bare strange star (SS) is different, since
the latter is self-bound and has a large (and roughly constant) density
and pressure throughout the star and right up to the surface \citep{Alford:2019oge}.
In this case the reduced mass per particle $\mu_{\mathrm{SS}}$ would
not be determined by low density nuclear physics but would be dependent
on the poorly constrained equation of state of such strange quark
matter. Moreover, this parameter should be significantly lower $\left(m-\mu_{\mathrm{SS}}\right)/m=O\!\left(1\right),$
which---in addition to the very different heat insulation---could
allow to discriminate or rule out such a scenario.

Since we neglected radiative losses all the expressions derived in
section \ref{sec:Accretion-heating} present upper bounds for the
total rotational $E_{\mathrm{rot}}\leq E^{\left(\mathrm{max}\right)}_{\mathrm{rot}}=\left(M_{\mathrm{acc}}/m\right)\Delta E_{\mathrm{rot}}$
and heat energy $E_{\mathrm{heat}}\leq E^{\left(\mathrm{max}\right)}_{\mathrm{heat}}=\left(M_{\mathrm{acc}}/m\right)\Delta E_{\mathrm{heat}}$
gained when accreting a large mass $M_{\mathrm{acc}}$ in an actual
accretion process. Although mass ejection and in particular the acceleration
of cosmic rays in the vicinity of compact objects is astrophysically
very important \citep{Blumer:2009jrd}, we don't expect that a significant
fraction of the initial gravitational energy is emitted this way.
Similarly the inspiralling charges clearly emit synchrotron emission.
However, most of the radiation is emitted close to the source were
the Lorentz factor is the largest, and it is correspondingly strongly
beamed towards instead of away from the source. In this region above
the magnetic poles based on the observed pulsed emission a dense plasma
can be expected, within which this synchrotron emission should be
reprocessed and eventually contribute to the overall observed hard
power law spectra \citep{Jaisawal:2024lps} from inverse Compton processes.
Therefore, the heating energy $E_{\mathrm{heat}}$ should be close
to the bound $E_{\mathrm{heat}}\approx E^{\left(\mathrm{max}\right)}_{\mathrm{heat}}$
and correspondingly the heating energy per particle $\Delta E_{\mathrm{heat}}$
should be close to the energy per particle of pulsed emission from
the compact source. However, the interaction of the magnetic field
with the disk as well as the spindown due to magnetic dipole radiation
\citep{1983bhwd.book.....S} can be important for the total torque
acting on the neutron star, so that it could be smaller (or maybe
even larger) than the spin-up energy per particle during accretion
$\Delta E_{\mathrm{rot}}$ suggests.

\begin{figure}
\includegraphics[scale=0.65]{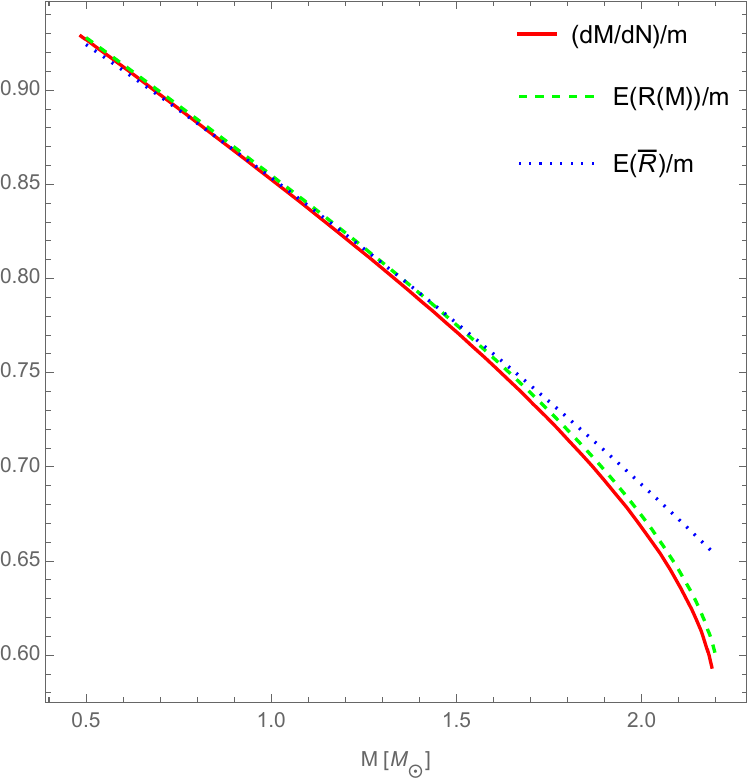}

\caption{\label{fig:star-energy-increase} The mass increase of a static neutron
star when adding another particle. The solid line shows the numerical
result obtained for an APR equation of state \citep{Akmal:1998cf},
the dashed line shows the EoS-independent analytic expression (but
using the corresponding mass-radius relation). The minor deviations
between these curves below the percent level can be attributed to
approximations in the EoS and numerical uncertainties. The dotted
curve shows the same when considering a mass-independent radius (determined
here by the turning point of the mass-radius curve which for the APR
equation of state is at $\bar{R}\approx11.5\,\mathrm{km}$).}
\end{figure}

For a neutron star the equation of state dependent form for the stellar
mass increase when adding another particle eq. (\ref{eq:mass-change})
(dashed curve) is in fig. \ref{fig:star-energy-increase} for a non-rotating
star compared to numerically performing the derivative $dM/dN$ for
an actual family of OV solutions for a static stellar configurations
(solid curve), when in both cases using the mass-radius relation for
the APR equation of state \citep{Akmal:1998cf}. As can be seen the
two indeed agree up to small numerical deviations below the percent
level, that can be attributed to approximations in the EoS and numerical
uncertainties. For most astrophysical sources their masses are not
known---let alone their radii. The same holds for the EoS of dense
matter that would relate them even in those cases where the masses
are known \citep{Ozel:2016oaf,Arf:2024lgp}. However, OV solutions
for the equilibrium structure of neutron stars predict that the radius
is very weakly dependent on the mass over a large range of realistic
values, which is for observed sources confirmed by NICER measurements
\citep{Choudhury:2024xbk,Salmi:2024aum}. Using the corresponding
characteristic radius (for comparison here $\bar{R}\approx11.5\,\mathrm{km}$
for the APR equation of state, which is consistent with the observationally
favored value \citep{Choudhury:2024xbk,Salmi:2024aum}) eq. (\ref{eq:mass-change})
provides away from the mass limit a remarkably good approximation
at the few percent level (dotted curve) that is completely independent
of the equation of state.

\begin{figure}
\includegraphics[scale=0.6]{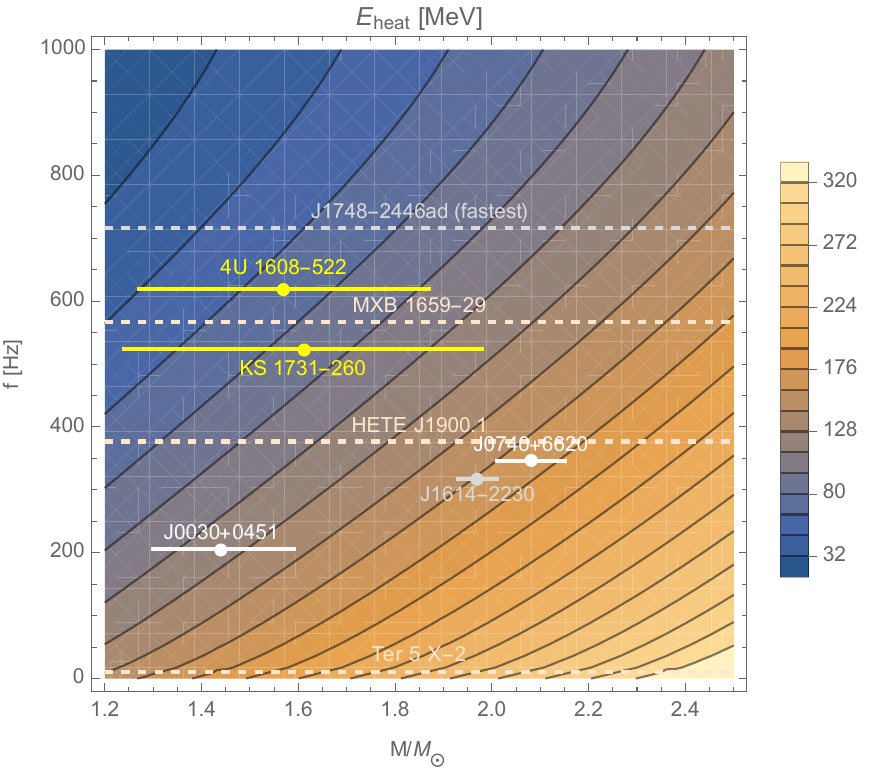}

\caption{\label{fig:heating-energy-corotation}The heating energy eq. (\ref{eq:heating-energy-corotation})
deposited by an accreted baryon falling from the co-rotation radius
onto a neutron star, compared to some exemplary LMXB sources, with
and without mass estimate, for which the rotation frequency is known
from burst oscillations \citep{Watts:2012kw,Potekhin:2019eya}.}
\end{figure}

The result for the heating per particle in accretion on a neutron
star is shown in fig. \ref{fig:heating-energy-corotation} for matter
falling from the co-rotation radius of a thin disk \citep{Shakura:1972te}.
As can be seen, faster spinning sources are heated significantly less.
The key reason for this is that the disk extends much closer to the
star so that the accreted energy reduces. The mass in contrast strongly
increases spacetime curvature so that the largest values for the heating
energy are expected for the heaviest sources. Depending on the spin
frequency the heating energy can change strongly so that the total
uncertainty if neither the mass nor the frequency is known can be
larger than an order of magnitude. Knowledge of the spin frequency
constrains the heating energy within roughly a factor three as is
shown for a few exemplary accreting sources in fig. \ref{fig:heating-energy-corotation}
(yellow dashed horizontal lines). As a reference also the fastest
known pulsar is shown (gray dashed line). If even the mass of the
source is known, as is shown for two so far rather poorly constrained
accreting sources (yellow), this reduces the uncertainty to roughly
a factor two. To estimate the improvement due to precise mass measurements,
we also show exemplarily a few well-constrained, non-accreting sources
observed by NICER (white) \citep{Miller:2019cac,Miller:2021qha} and
via Shapiro delay \citep{Demorest:2010bx} (gray), where the uncertainty
can become significantly smaller, namely down to the 10\% level.

\begin{figure}
\includegraphics[scale=0.6]{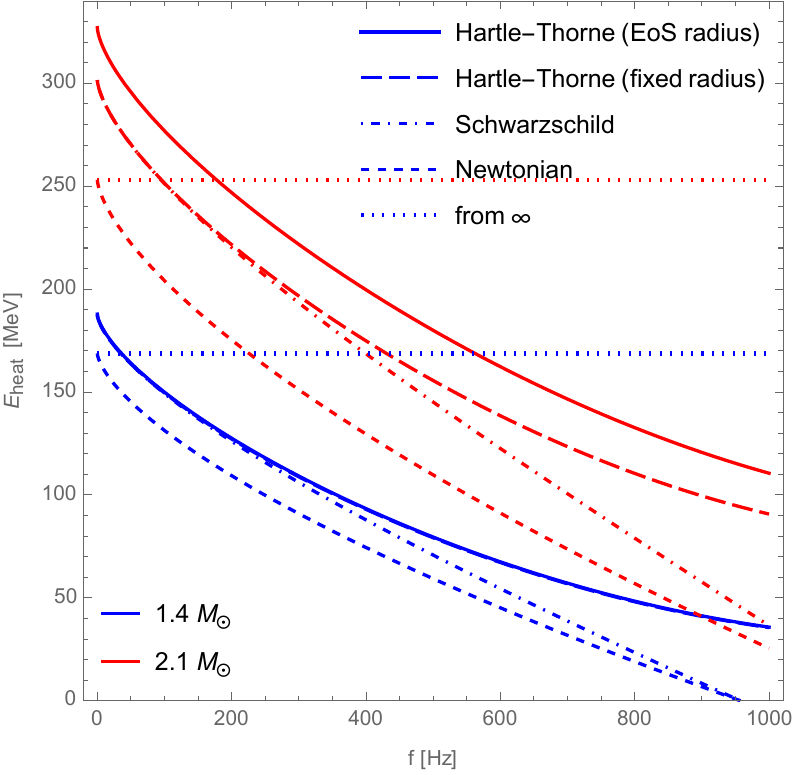}

\caption{\label{fig:heating-energy-approximations}The heating energy deposited
by an accreted baryon falling from the co-rotation radius in the different
approximations in eq. (\ref{eq:heating-energy-corotation}). The solid
and long-dashed lines give the full Hartle-Thorne expression when using
the mass-radius relation for an APR equation of state \citep{Akmal:1998cf}
respectively a fixed radius of $\bar{R}\approx11.5\,\mathrm{km}$
(defined by the turning point around which the radius is roughly mass-independent).
The other curves show the approximate expressions in eq. (\ref{eq:heating-energy-corotation}).}
\end{figure}

A comparison of the results in the different approximations is shown
in fig \ref{fig:heating-energy-approximations}. While the naive estimate
of the heating energy via the gravitational potential energy a particle
gains when it falls from infinity depends merely on the mass, the
other estimates also strongly depend on the frequency of the source.
As can be seen the results in any of the considered approximations
can deviate significantly from the naive estimate. 

\section{Conclusions}

In this work, we have considered the accretion onto compact stars
and determined the energy per particle that spins up and heats the
star. To this end we performed a controlled analysis within a slow
rotation expansion and showed that in presently observed neutron stars---up
to percent corrections---the Hartle-Thorne metric reduces to the
Lense-Thirring metric. I.e. although compact stars are much richer
systems than black holes, as far as the spacetime around them is concerned,
they in general do not have much hair either---one might say they
have ``eyebrows'' but are otherwise completely bald. This applies
to any process in the vicinity of a neutron star and correspondingly
this result goes beyond the scope of the present work. 

This framework allowed us to derive explicit analytic expressions
for the accreted energies based on general conservation equations.
While the problem at first sight seems to depend on the equation of
state of dense matter, we show that it is actually effectively independent
of this poorly constrained relation. We find that the most important
effects are that for a thin, magnetized disk the matter merely falls
from the inner edge of the accretion disk and already carries rotational
kinetic energy. These two effects are of similar size in line with
the virial theorem and can have an impact of nearly an order of magnitude.
Also relativistic corrections are important and can amount to several
tens of percent. In contrast the effect of frame dragging and other
general-relativistic rotational corrections are merely at the percent
level. Although frame dragging is linear in the spin rate in the Hartle-Thorne
metric, its contributions to the spin-up and heating energy are in
the co-rotating case only quadratic. Therefore, while frame dragging
effects can indeed be sizable close to the star, in a magnetized disk
where all charges, i.e. atomic nuclei and electrons, are forced to
co-rotate with the star, the co-rotation radius is significantly further
away than the ISCO, where frame dragging effects are already minor.
The case of a hot thick disk, where the geometry strongly changes
requires a separate analysis, and as our results show, it can be expected
that frame dragging effects play a more important role in this case.
An interesting aspect in this regard is that if there is a neutral
matter component, a (dilute) neutral part of the disk could even for
a thin disk extend all the way to the ISCO and therefore significantly
closer to the compact object. While neutrons could in principle be
created due to electron capture processes or dissociation of light
nuclei (e.g. due to reactions with the wind of the compact object)
their lifetime is likely far too short to constitute a significant
neutral extension of the disk, and this would require more exotic,
e.g. dark matter, particles.

Our results are relevant for accreting x-ray binaries since the heating
energy powers their strong x-ray emission. It is relevant for the
dynamics of the accretion disk that can exhibit instabilities \citep{1981AA...104L..10M},
thought to be responsible for the transient nature of the accretion
when taking into account the effect of irradiation by the compact
object \citep{1998MNRAS.293L..42K,Ertan:2015nna,Ertan:2020uos}. In
particular, these transient systems have the potential to probe the
interior composition of the compact star via its neutrino cooling
in a long term thermal steady state \citep{Brown1998,Yakovlev:2002ti,Arf:2024lgp}
and the heating energy is required to determine the average mass accretion
rate. Our results show that without knowledge of both the spin rate
and the mass of the compact source the accretion energy cannot be
narrowly constrained. While the spin frequency can be determined from
burst oscillations \citep{Watts:2012kw} seen in part of the observed
sources, the masses are still mostly unknown. This introduces significant
uncertainties that the derived expressions will allow us to consistently
estimate and take into account.
\begin{acknowledgments}
It is a pleasure to thank Mehmet Ali Alpar, Yavuz Eksi and Tolga Güver
for very helpful discussions. This work was supported by the Turkish
Research Council (TÜBITAK) via project 125F016.
\end{acknowledgments}

%

\end{document}